\documentclass[english,prb,twocolumn,showpacs,floatfix,superscriptaddress]{revtex4}

\usepackage{float}
\usepackage[latin1]{inputenc}
\usepackage{graphicx}
\usepackage{hyperref}

\usepackage{color}
\usepackage{changebar}

\begin{document}
\newcommand{\ldg}{electron Land{\'e}-g-factor}
\newcommand{\tcr}[1]{\textcolor{red}{#1}}
\newcommand{\vbt}[1]{\verb"#1"}
\newcommand{\hidden}[1]{}

\preprint{R{\"o}mer}

\title{Electron spin relaxation in bulk GaAs for doping densities close to the metal-to-insulator transition}

%
%





\author{M. R{\"o}mer}
\email{roemer@nano.uni-hannover.de}
\homepage{http://www.nano.uni-hannover.de/}
\author{H. Bernien}
\author{G. M{\"u}ller}
\affiliation{Institute for Solid State Physics, Gottfried Wilhelm
Leibniz University of Hannover, Appelstr. 2, 30167 Hannover,
Germany}
\author{D. Schuh}
\affiliation{Institute for Experimental and Applied Physics, Regensburg University, D-93040 Regensburg, Germany}
\author{J. H{\"u}bner}
\affiliation{Institute for Solid State Physics, Gottfried Wilhelm
Leibniz University of Hannover, Appelstr. 2, 30167 Hannover,
Germany}
\author{M. Oestreich}
\affiliation{Institute for Solid State Physics, Gottfried Wilhelm
Leibniz University of Hannover, Appelstr. 2, 30167 Hannover,
Germany}
\affiliation{Centre for Quantum Engineering and Space-Time Research (QUEST), Hannover, Germany}

\date{\today}

\begin{abstract}
We have measured the electron spin relaxation rate and the
integrated spin noise power in n-doped GaAs for temperatures
between 4~K and 80~K and for doping concentrations ranging from
$2.7\times 10^{-15}~\text{cm}^{-3}$ to $8.8\times 10^{-16}~
\text{cm}^{-3}$ using spin noise spectroscopy. The temperature
dependent measurements show a clear transition from localized to
free electrons for the lower doped samples and confirm mainly free
electrons at all temperatures for the highest doped sample. While
the sample at the metal-insulator-transition shows the longest spin relaxation time at low
temperatures, a clear crossing of
the spin relaxation rates is observed at 70~K and the highest
doped sample reveals the longest spin relaxation time above 70~K.
\end{abstract}

\maketitle

\section{Introduction}

Non-equilibrium electron spin relaxation in semiconductors has
been studied  for nearly forty years and is today in the focus of
intense research due to the vision of semiconductor spintronic
devices. The most prominent model system in this field is n-doped
bulk GaAs since high quality GaAs is easily available and spin
polarized electrons can be efficiently excited and detected by
circularly polarized light excitation and photoluminescence
detection. An excellent overview about the most
relevant spin relaxation mechanisms and experimental results in
this material system is given in Refs.~\onlinecite{Meier1984,DzhioevPRB2002,Dyakonov2008}. Presently the non-equilibrium electron spin relaxation in n-doped GaAs is perfectly understood for low temperatures and doping concentrations well above the metal-to-insulator transition
(MIT) where the Dyakonov-Perel (DP) spin relaxation dominates. The
same is true for all doping concentrations at high temperatures
\cite{oertel:132112} where all electrons are delocalized and the spin
dynamics can be well described by semiconductor spin Bloch
equations \cite{jiang:125206}. The situation becomes much more complex for the doping regime at
the metal-to-insulator transition at low temperatures where the
spin relaxation times turn out to be very long. In this regime,
the theoretical description becomes more complicated due to the
intricate interplay of localized and free electrons. At the same
time, the experiments become more difficult since optical
excitation changes momentum scattering times and the ratio of free
and localized electrons.

The most detailed work on the non-equilibrium, low temperature
electron spin relaxation time in dependence on doping density has
been performed by Dzhioev et al.\cite{DzhioevPRB2002}. They identified hyperfine
interaction, anisotropic exchange interaction, and the DP
mechanism as the dominant spin relaxation mechanisms for doping
densities below $2\times 10^{15}~\mathrm{cm}^{-3}$, between $2\times
10^{15}~\mathrm{cm}^{-3}$ and the MIT, and above the MIT,
respectively. Dzhioev et al. have measured the non-equilibrium
spin relaxation by optical Hanle depolarization experiments and consequently observed
a strong dependence of the spin relaxation time on excitation
power. In this publication, we complement their experiments and
measure the {\em equilibrium} spin relaxation time by
nearly perturbation free spin noise spectroscopy for different doping
concentrations over a wide temperature range.

\section{Samples and Experimental Setup}

The samples are one 2~$\mu$m thick, silicon-doped, molecular-beam
epitaxy (MBE) grown GaAs sample with AlGaAs barriers with a doping concentration of $n_D
=1\times 10^{14}~\mathrm{cm}^{-3}$ (sample A$^{\mathrm{vl}}$; very
low doped) and three silicon-doped, Chochralskii-grown, GaAs
wafers with doping concentrations of
$n_D = 2.7\times 10^{15}~\mathrm{cm}^{-3}$ (sample
B$^{\mathrm{low}}$; low doped), $n_D = 1.8\times
10^{16}~\mathrm{cm}^{-3}$ (sample C$^{\mathrm{MIT}}$; doping close
to the MIT), and $n_D = 8.8\times 10^{16}~\mathrm{cm}^{-3}$ (sample
D$^{\mathrm{high}}$; high doping concentration). All samples are equipped with a high quality, $\lambda/4$ silicon nitride antireflection coating. For transmission experiments, the MBE grown GaAs/AlGaAs layer has been lift-off from the GaAs substrate and Van-der-Waals bonded to a sapphire substrate.

\begin{figure}[tbp]
    \centering
        \includegraphics[width=0.45\textwidth]{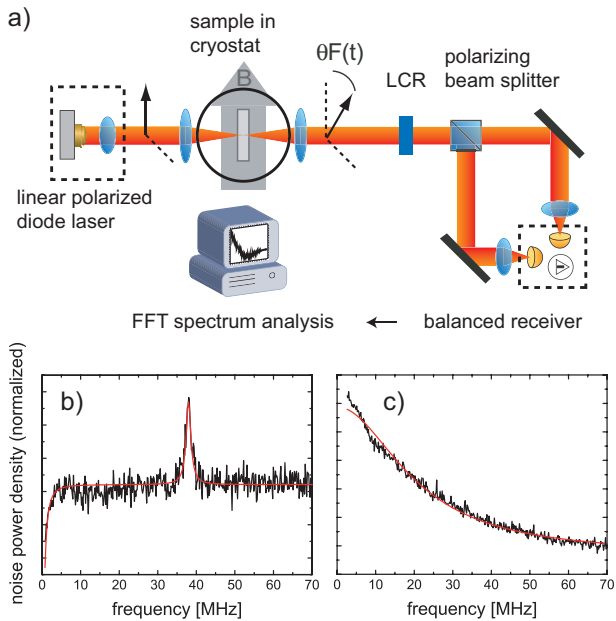}
    \caption{(a) Spin noise spectroscopy setup.
    The external magnetic field shifts the center frequency
    of the spin noise signal by the electron Larmor frequency. (b) Typical
    spin noise spectrum acquired by subtracting spin noise spectra measured
    at B=6 and 0~mT for sample C$^{\mathrm{MIT}}$ at 4~K. (c)
    Typical spin noise difference spectrum acquired by selective suppression
    of spin noise with a liquid crystal retarder for sample B$^{\mathrm{low}}$
    at 10~K.}
    \label{fig:setup}
\end{figure}

Figure~\ref{fig:setup}(a) depicts the experimental spin noise
spectroscopy setup.\cite{RoemerRSI2007} The light source is a low
noise, tunable diode laser in Littmann configuration. A Faraday
isolator avoids feedback into the laser and a single mode fibre is
used as spatial filter to ensure a Gaussian spatial laser profile.
The laser light is focused to a beam waist of $w_0=80~\mu$m in the sample which is mounted in a He cold finger
cryostat. The wavelength of the linearly polarized laser light is
tuned below the GaAs band gap to avoid laser light absorption. The
GaAs donor electrons are in thermal equilibrium with zero mean
spin polarization but the temporal statistical fluctuations of the
spin polarization are unequal to zero and yield a fluctuating Faraday rotation
of the linear laser light polarization. This Faraday rotation is
measured by a combination of a polarizing beam splitter and an 80~MHz balanced photoreceiver. The electrical signal is
amplified by a low noise amplifier and passed through a 67~MHz low
pass filter. The fluctuation signal is digitized with 180~MS/s and
16~bit in the time domain and Fourier transformed in real time. To
eliminate the white photon shot noise in the measured spectra, a
second spin noise spectrum with the spin noise either
shifted in frequency or totally suppressed is acquired and
subtracted. Figure~\ref{fig:setup}(b) shows a typical difference
spectrum obtained by subtracting two spectra acquired at $B=6$ and 0~mT, respectively, i.e., the second of the two spin noise spectra has been shifted by the electron Larmor frequency.
Figure~\ref{fig:setup}(c) shows a typical difference spectrum
acquired by selective spin noise suppression. The selective
suppression of the noise signal is performed by a liquid crystal
retarder (LCR) after the cryostat with the fast axis aligned parallel to the
polarization plane of the laser. The LCR can be set to retardance of
$\lambda /4$ (spin noise suppressed) or $\lambda /2$ (no change).
When possible both methods are combined in a double difference
technique to eliminate remaining offsets, e.g., due to a slightly different
transmission of the LCR for $\lambda/2$ and $\lambda/4$
retardation.

To assess the influence of unwanted residual optical
excitation or time-of-flight broadening on the measured spin
relaxation time, the laser spot diameter has been varied from
$30\,\mu$m to $120\,\mu$m while keeping the total laser power constant.
Enlarging the focus spot reduces the power density in the sample
and time-of-flight broadening, i.e., minimizes spurious effects due to the measurement technique\cite{muller:206601}. The interrelation of the energy position of the probe light relative to the
electronic resonance has been studied
previously\cite{RoemerRSI2007}.
The optimum results have been obtained for a probe laser wavelength
of 840 nm, 4~mW power, and 80 $\mu$m spot size. Longer
wavelengths, lower laser power, or larger spot diameters did not change
the measured spin relaxation times but reduced the signal to
noise ratio significantly thus increasing the required averaging
time.
Different measurement conditions have been used for sample A$^{\mathrm{vl}}$ as the
short spin lifetime and the low donor concentration lead to a small noise signal at large laser detuning.
The sharp excitonic lines at low temperatures and low dopand concentration allow
the use of a laser wavelength of 820.7 nm with only negligible absorption.

\section{Experimental Results}
In the following we present the temperature dependence of the spin
relaxation rate $\Gamma_s$ and of the spin noise power $P_s$ and
discuss the influence of free and localized electrons on the spin
relaxation. The samples can be classified into three categories by their doping concentration namely the completely localized phase (sample A$^\mathrm{vl}$), the metallic phase (sample D$^{\mathrm{high}}$), and the mixed phase around the MIT (samples B$^{\mathrm{low}}$ and C$^{\mathrm{MIT}}$). Figure~\ref{fig:temp-dep}(a) shows the temperature
dependence of the spin relaxation rate $\Gamma_s$ for sample
B$^\mathrm{low}$, C$^\mathrm{MIT}$, and D$^\mathrm{high}$ in the
temperature range from 4~K to 80~K and for sample A$^\mathrm{vl}$ at 8~K.

\subsection{Metallic phase}

\begin{figure}[tbp]
    \centering
        \includegraphics[width=0.45\textwidth]{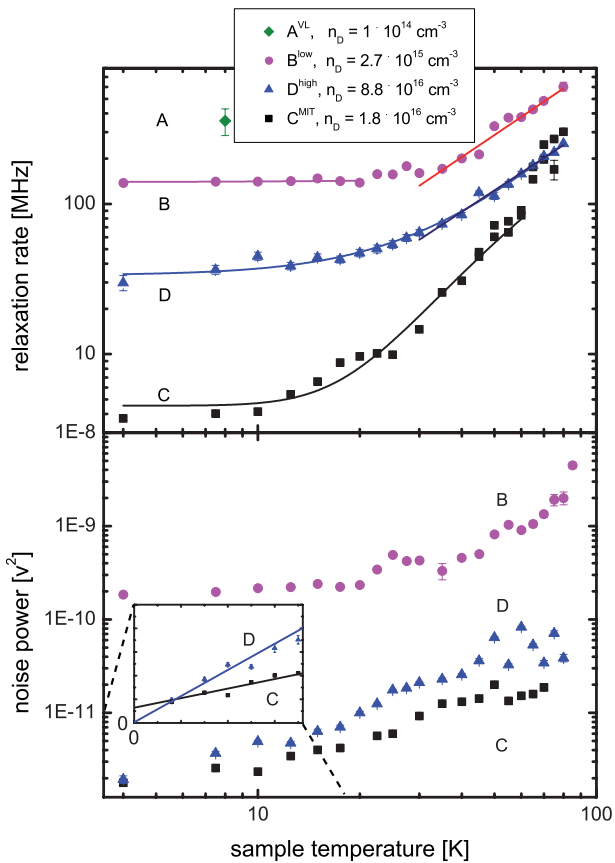}
    \caption{(a) Temperature depended measurement of the spin
    relaxation rate for the samples B-D using a large focus and a large detuning of
    the laser wavelength from the optical transitions (probe laser wavelength
$\lambda=$~840 nm) and the spin relaxation rate for sample A at 8~K. The solid lines are fits to the data. The fit parameters are listed in table \ref{table:Gamma}. (b) Temperature dependence of the integrated
spin noise power for samples B-D.
 The noise powers have been scaled to account for the different thicknesses of the samples.
 The inset shows the low temperature noise power for samples C and D on a linear scale.}
    \label{fig:temp-dep}
\end{figure}

The doping of the highest
doped sample D$^\mathrm{high}$ is well above the MIT and the
conduction band is populated even at very low temperatures due to
the hybridization of impurity and conduction band, i.e., all
electrons are delocalized, the Fermi level is at low temperatures
well in the conduction band, the dominant spin relaxation
mechanism is the DP mechanism, and ionized impurity scattering is
the main electron scattering mechanism. The delocalization is
substantiated by the temperature dependence of the integrated spin
noise power (see Fig.~\ref{fig:temp-dep}(b)) which extrapolates to
zero noise power at zero temperature due to spin Pauli blockade.
In the range from 30~K to 80~K the temperature dependence is
proportional to $T^{1.48 \pm0.06}$ which is in very good agreement with the expected $T^{3/2}$
dependence in the case of charged impurity scattering\cite{Zutic2004}.
Also the low temperature value of $\Gamma_s$ measured by SNS is in
rather good agreement \footnote{The discrepancy \hidden{by 25 percent is
within the error bars of the SNS and the Hanle measurements but}
might result from the higher impurity concentration and a
resulting faster momentum scattering time in our
Chochralskii-grown GaAs. \hidden{soll das wirklich als within error
bars geschrieben werden bei mehr 25 \% Abweichung?}} with the
Hanle measurement by Dzhioev et al. \cite{DzhioevPRB2002}. Such an
agreement is expected since the influence of the weak optical
excitation in the Hanle experiment is small for this doping concentration.

\begin{table*}[htb]
  \centering
  \begin{ruledtabular}
  \begin{tabular}{l l l}
  sample & spin relaxation rate & temperature regime\\
  \hline
  A$^{\mathrm{vl}}$ & $\Gamma_s = 357 \pm 72$~MHz & T = 8~K\\
  B$^{\mathrm{low}}$ & $\Gamma_s = 140 \pm 2$~MHz & T = 4..20~K\\
  B$^{\mathrm{low}}$ & $\Gamma_s \propto T^{1.48 \pm 0.09}$ & T = 30..80~K\\
  C$^{\mathrm{MIT}}$ & $\Gamma_s = 3065~\mathrm{MHz} \times \mathrm{exp}(-(1028~\mathrm{K}/T)^{1/2})+4.5~\mathrm{MHz}$ & T = 4..60~K\\
  D$^{\mathrm{high}}$ & $\Gamma_s \propto  T^{1.48 \pm 0.06}$ & T = 30..80~K\\
  D$^{\mathrm{high}}$ & $\Gamma_s = 0.04~\mathrm{MHz} \times (T/\mathrm{K})^{1.96 \pm 0.08} + 33.5~\mathrm{MHz}$ & T = 2..80~K
  \end{tabular}
  \end{ruledtabular}
  \caption{Temperature dependence of the spin relaxation rates
  from fits to $\Gamma_s$ from Fig.~\protect\ref{fig:temp-dep}(a).}
  \label{table:Gamma}
\end{table*}

\subsection{Non-interacting donors}
\begin{figure}[htbp]
    \centering
        \includegraphics[width=0.45\textwidth]{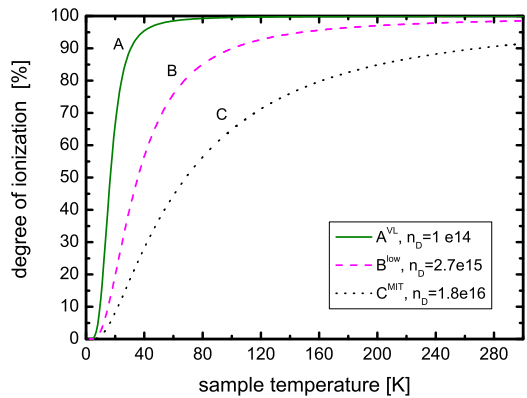}
    \caption{Temperature dependence of the degree of electron ionization
    for the three lowest doped samples calculated by Blakemore's
    equation\cite{blakemore:2002}. The validity of Blakemore's equation
    decreases with increasing doping concentration due to the formation
    of impurity bands, i.e., the calculation for sample~C is only of
    limited significance.}
    \label{fig:fig-blakemore}
\end{figure}

The other extreme concerning doping is sample A$^\mathrm{vl}$
where the average distance between two donors is more than 200~nm.
The electrons are at low temperatures completely localized at the
donor atoms and electron-electron interaction can be neglected in
good approximation. The dominant spin relaxation is in this case hyperfine interaction with nuclear spins which can be expressed by
an effective nuclear magnetic field $B_N$. The strength and direction of $B_N$ varies at thermodynamic equilibrium stochastically from donor to donor resulting in a dephasing of the ensemble electron spin polarization due to an inhomogeneous Larmor precession. The measured spin relaxation time in sample A$^\mathrm{vl}$ is $2.8 (\pm 0.7)$~ns (8~K\footnote{Sample A$^{\mathrm{vl}}$ has been measured in a helium free cryostat, where 8~K was the minimum temperature.}) which is in fair agreement with the calculated value of 3.6~ns by  Merkulov et al. \cite{merkulov:205309}. Following the calculations of Merkulov et al. one third of the electrons, i.e., electrons with spins aligned along $B_N$, show orders of magnitudes longer spin relaxation times when no external magnetic field is applied. In the thin sample A$^\mathrm{vl}$ we do not observe such long $\tau_s$ since the free exciton line is broadened due to strain and the resulting residual absorption prevents the detection of such very long spin dephasing times. The free (bound exciton) resonance is inhomogeneously broadened by 0.9~meV (3~meV) due to inhomogeneous strain in the lift-off sample which results in residual light absorption. The residual light absorption can not be avoided in this sample by longer laser wavelengths due to a scarce signal-to-noise ratio. The light absorption leads to the creation of trions which suppress spin noise by Pauli blockade, i.e., the by SNS measured $\tau_s$ is a lower bound. In contrast, the optical non-resonant excitation in Hanle experiments yields free carriers which lead to an increased averaging of the the hyperfine interaction and hence a decrease of the spin relaxation rate \cite{Furis2006}, i.e., Hanle experiments give in first approximation an upper bound for $\tau_s$. However, the intricate density dependence at very low densities complicates the interpretation of Hanle experiments and a linear extrapolation of $\tau_s$ to zero excitation is ambiguous.
The spin relaxation time $\tau_s$ measured by SNS is by
a factor of two smaller than the value measured by Dzhioev et al. by the Hanle effect \cite{DzhioevPRB2002}, which we attribute to the reasoning pointed out above.
The large relative error bar of $\tau_s$ for sample A$^\mathrm{vl}$ compared to the other samples
is due to the fact that the spin noise spectrum is broader than the detection bandwidth of the setup which complicates the data analysis. We have not measured the temperature dependence of $\tau_s$ for this sample since ionization of the donor bound electrons increases significantly above 10~K (see Fig.~\ref{fig:fig-blakemore}). 

\subsection{Mixed phase}

The height of the localization potential and degree
of localization for samples close to the metal insulator transition delicately depend on the donor concentration and temperature. For lower doping concentrations the localization potential is higher, electron-electron interaction is weaker, and delocalization occurs besides the higher localization potential at lower temperatures. \hidden{see Furis et al. \cite{Furis2006} with $\tau_{s}=600$~ns at 10~K for $n=4\times 10^{15}\text{cm}^{-3}$!!}

Figure~\ref{fig:temp-dep} shows inter alia the temperature
dependence of $\Gamma_s$ and $P_s$ for the second lowest doped sample
B$^{\mathrm{low}}$. Both are approximately
constant for temperatures up to 20~K which is consistent with
localized electrons. At low temperature $\tau_s$ is longer than
in sample A$^{\mathrm{vl}}$ which results from the exchange interaction
between the donor electrons,\cite{kavokin:114009} i.e., the spin interaction causes an effective averaging over the locally different nuclear fields which increases $\tau_s$. Interestingly,
$\tau_s=7 \, \mbox{ns}$ measured by SNS is significantly shorter than $\tau_s$ measured by Hanle experiments where $\tau_s$ varies for comparable doping concentrations between $\sim 26$~ns (Colton
et. al. \cite{colton:pss2002}) and $\sim 180$~ns (Dzhioev et al.\cite{DzhioevPRB2002}). This large spread of the values of $\tau_s$ can only partially be explained by the strong dependence
of $\tau_s$ on the exact doping concentration, the ratio of compensation, and the purity of the individual sample. Most probably, the longer spin relaxation times observed via the Hanle
effect are again at least to some extend -- as for doping concentrations of $10^{14}$~cm$^{-3}$ -- due to optical excitation\cite{PhysRevB.70.113201}.

At this point, we want to overcome a recent misleading statement against SNS which results from a very commendable and detailed comparison of Hanle and SNS by Crooker et al. \cite{crooker:035208}. The authors state that SNS is by no means a panacea and that accurate measurements of $\tau_s$ are more readily and quickly obtained using conventional techniques based on the Hanle effect, which is particulary correct for higher doping concentrations and elevated temperatures. However, Figure~5 of Ref.~\onlinecite{crooker:035208} implies to non-experts that the Hanle
technique is superior to SNS and less perturbative, which does not hold in general. As stated in Ref. \onlinecite{crooker:035208}, SNS is nonperturbative for long wavelengths, however the shorter spin relaxation time results from time-of-flight broadening which is known from experiments on atomic gases \cite{CrookerNATURE2004} and has recently been verified for the first
time in semiconductor quantum wells \cite{muller:206601}. This effect is absent for localized carriers and can be overcome by a sufficiently large laser spot for free carriers.

Figure~\ref{fig:temp-dep} also shows for sample B$^{\mathrm{low}}$ an
increase of $\Gamma_s$ for temperatures above 30~K which is
proportional to $T^{1.48 \pm 0.09}$. This increase of $\Gamma_s$ results
from ionization of the electrons and efficient spin relaxation of
the ionized electron spins due to the DP spin relaxation
mechanism. According to Blakemore's formula, the degree of ionized
electrons at 30~K is about 50~\%\ and the percentage rises by 1.5~\%\ per Kelvin
at this doping concentration, i.e., spin relaxation by nuclear
hyperfine interaction can be neglected above 30~K since the efficiency of DP
increases with temperature and the efficiency of spin relaxation
by hyperfine interaction decreases due to the interaction of
localized and free electrons and the resulting efficient averaging
over the nuclear fields.

\subsection{Close to the metal-to-insulator transition}

Sample C$^{\mathrm{MIT}}$ has a donor concentration right at the
MIT which occurs for GaAs at a concentration of $n_D \approx 2
\times 10^{16}~\mathrm{cm}^{-3}$. The MIT is characterized by overlapping donor
atoms which start to form an impurity band below the conduction
band\cite{AbramAIP1978} and allows electrical conductivity over
macroscopic distances by percolation paths, i.e., the low
temperature Fermi level is located inside the impurity band
\cite{sss45}, such that some electrons are completely delocalized but
most electrons are confined on a macroscopic scale but therein
delocalized. Figure~\ref{fig:temp-dep}(b) shows that $P_s$ decreases for decreasing temperature but does not completely vanish for extrapolation to zero temperature which substantiates partial
Pauli spin blockade. The overall spin noise power is smaller for sample C$^{\mathrm{MIT}}$ compared to sample D$^{\mathrm{high}}$ due to the lower electron density in sample C$^{\mathrm{MIT}}$ and is also smaller compared
to the lower doped sample B$^{\mathrm{low}}$ due to strong excitonic
enhancement of the spin noise signal in sample B$^{\mathrm{low}}$ even at
moderate temperatures.

The squares in Fig.~\ref{fig:temp-dep}(a) depict the temperature
dependence of $\Gamma_s$ for sample C$^{\mathrm{MIT}}$. We observe a long spin relaxation time
of 267~ns at 4 K which is about 200~ns longer than the value
reported in Ref.\,\onlinecite{DzhioevPRB2002} for a similar sample.
The sample has the longest $\tau_s$ at low
temperatures but $\Gamma_s$ increases strongly for temperatures
above 10~K. Above 70~K, $\Gamma_s$ of sample C$^{\mathrm{MIT}}$ becomes larger than
$\Gamma_s$ of the highly doped sample D$^{\mathrm{high}}$, i.e., samples at
the metal-to-insulator transition have only the longest spin
relaxation times at low temperatures but not at high temperatures.
This crossing of $\Gamma_s$ at finite temperatures is not
surprising since the Fermi distribution becomes a
Maxwell-Boltzmann distribution and hence the average electron
energy becomes independent of doping concentration. At the same
time, electron-electron and electron-impurity scattering are
faster for higher doping concentrations which leads to a more
efficient motional narrowing and less efficient spin relaxation by
the DP mechanism. Simple estimations of $\Gamma_s$ by the DP
mechanism using the Brooks-Herring approach for charged impurity
scattering confirm theoretically the crossover of $\Gamma_s$ at
$T\approx 70$~K for sample C$^{\mathrm{MIT}}$ and D$^{\mathrm{high}}$. We estimate
that for free electrons and carrier temperatures above the Fermi
temperature $(E_{F}/k_{B})$ the sample with the highest doping concentration shows
the longest spin dephasing time.

\begin{figure}[htbp]
    \centering
        \includegraphics[width=0.5\textwidth]{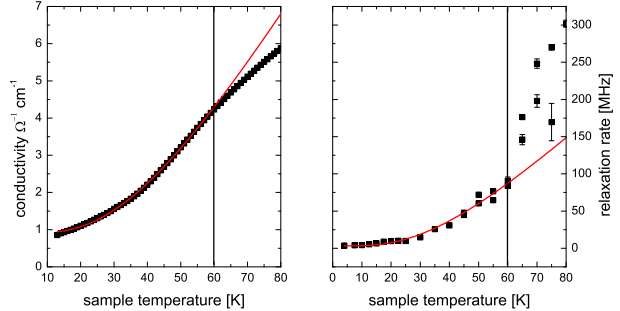}
    \caption{Temperature dependence of the spin relaxation in the
    hopping regime depends on the conductivity. Hopping transport
    describes the temperature dependence well for temperatures below 60 K.}
    \label{fig:fig-leit}
\end{figure}

Up to temperatures of 60~K, $\Gamma_s$ of sample C$^{\mathrm{MIT}}$ does
not follow the usual temperature dependence of the DP spin
relaxation mechanism calculated by scattering of free electrons at
charged impurities. In fact, we have measured the conductivity
$\sigma$ versus temperature which is up to a temperature of 60~K
well described by hopping transport with $\sigma=\sigma_0
\mathrm{exp}(-(T_0/T)^{1/2})+\sigma_m$ as shown in Fig.~\ref{fig:fig-leit}(a). The exponential term describes the hopping transport and $\sigma_m$ is a metal-like contribution to the concuctivity\cite{PhysRevB.36.4748}. For
temperatures higher than 60~K, most electrons are delocalized and
the sample shows normal metallic conductivity. In Fig.~\ref{fig:fig-leit}(b))
a fit with the same temperature dependence ($\Gamma_s = \Gamma_0\mathrm{exp}(-(T_0/T)^{1/2})+\Gamma_m$) and the same value for $T_0$ has been applied to the spin relaxation rate
showing that $\Gamma_s$ is proportional to the conductivity.

Spin relaxation in the hopping regime can occur via a DP like spin
relaxation mechanism as described by
Shklovskii\cite{shklovskii:193201,kavokin:114009} or via
anisotropic spin exchange (Dzyaloshinskii-Moriya\cite{PhysRev.120.91,Dzyaloshinsky1958241} interaction
(DM)) as described by Kavokin\cite{PhysRevB.64.075305}. Putikka et~al.\cite{PhysRevB.70.113201} estimate the spin relaxation rate due to the DM mechanism by
\begin{equation}
\Gamma_{DM}=1/\tau_{DM}=\alpha_{DM} N_D a^3_B f_{DM}(T),
\end{equation}
where $a_B=10.6$~nm is the Bohr radius of the bound electron,
$\alpha_{DM}(\mbox{th})=0.01 \, \mbox{ns}^{-1}$ (theoretical
value) or $\alpha_{DM}(\mbox{exp})=0.03 \, \mbox{ns}^{-1}$
(experimental value) is a constant relaxation rate, and
$f_{DM}(T) \approx 32$ at $T=5$ K is a weakly
temperature-dependent function. Accordingly, we calculate for
sample C$^{\mathrm{MIT}}$ at 5~K a spin relaxation time in the
range of $\tau_{DM}(\mbox{th}) \approx 150$~ns and
$\tau_{DM}(\mbox{exp}) \approx 485$~ns, which is consistent
with the measured value of 267~ns.

\section{Summary}

The temperature dependence of the electron spin relaxation time $\tau_s$ in bulk
n-GaAs has been studied using spin noise spectroscopy. 
The results are summarized in Table \ref{table:Gamma}.
The different samples with doping concentrations in the vicinity of the
metal-to-insulator transition cover the range from fully localized
to entirely free electrons which is confirmed by temperature
dependent measurements of the spin noise power. At high
temperatures, all measurements are consistent with DP spin
relaxation of free electrons. At low temperatures and low doping
concentrations, $\tau_s$ is in good approximation independent of
temperature since the electrons are localized. At the same time, $\tau_s$
measured by spin noise spectroscopy is shorter than in comparable
measurements by the Hanle technique which can be attributed to
weaker perturbation in the case of SNS. For doping densities at
the metal-to-insulator transition and temperatures up to 60~K,
both conductivity and $\tau_s$ are well described by hopping
transport. Interestingly, the low temperature spin relaxation time
is longest for doping densities at the metal-to-insulator
transition but $\Gamma_s$ increases with temperature less rapidly
for higher doped samples. At 70~K, a crossing of $\Gamma_s$
appears for the two doping concentrations $n_D = 1.8\times
10^{16}~\mathrm{cm}^{-3}$ and $n_D = 8.8\times
10^{16}~\mathrm{cm}^{-3}$ and $\tau_s$ of the higher doped sample
becomes longer than $\tau_s$ for the lower doped sample. This
crossing results mainly from the transition from Fermi-Dirac to
Maxwell-Boltzmann electron distribution and the faster electron
momentum scattering by charged impurities in the higher doped
sample.

\section{Acknowledgement}

This work was supported by the German Science Foundation (DFG
priority program 1285 `Semiconductor Spintronics'), the Federal
Ministry for Education and Research (BMBF NanoQuit), and the
Centre for Quantum Engineering and Space-Time Research in Hannover
(QUEST). G.M.M. acknowledges support from the Evangelisches
Studienwerk.


\clearpage

\end{document}